\documentclass[reprint,showpacs,superscriptaddress,floatfix,twocolumn]{revtex4-2}

\usepackage{epsfig}
\usepackage{amsmath}
\usepackage{amssymb}
\usepackage{amsfonts}
\usepackage{mathptmx}
\usepackage{dcolumn}
\usepackage{eucal}
\usepackage{bm}
\usepackage{color}
\usepackage[colorlinks,linkcolor=blue,citecolor=blue]{hyperref}
\usepackage{epstopdf}
\usepackage{url}
\usepackage{graphicx,setspace}

\usepackage{dsfont}

\def\be{\begin{equation}}
\def\ee{\end{equation}}
\def\bea{\begin{eqnarray}}
\def\eea{\end{eqnarray}}
\def\ba{\begin{array}{l l}}
\def\ea{\end{array}}

\begin{document}                  

\title{Magnetic exchange interaction in spin-valve with chiral spin-triplet superconductor}

\author{Alfonso Romano}
\affiliation{Dipartimento di Fisica ‘E.R. Caianiello’, Universit\'a di Salerno, Fisciano, Salerno I-84084, Italy}
\affiliation{CNR-SPIN, c/o Universit\'a di Salerno- Via Giovanni Paolo II, 132 - 84084 - Fisciano (SA), Italy}
\author{Canio Noce}
\affiliation{Dipartimento di Fisica ‘E.R. Caianiello’, Universit\'a di Salerno, Fisciano, Salerno I-84084, Italy}
\affiliation{CNR-SPIN, c/o Universit\'a di Salerno- Via Giovanni Paolo II, 132 - 84084 - Fisciano (SA), Italy}
\author{Mario Cuoco}
\affiliation{CNR-SPIN, c/o Universit\'a di Salerno- Via Giovanni Paolo II, 132 - 84084 - Fisciano (SA), Italy}
\affiliation{Dipartimento di Fisica ‘E.R. Caianiello’, Universit\'a di Salerno, Fisciano, Salerno I-84084, Italy}

\begin{abstract}
The coupling between two ferromagnets separated by a superconductor has been mostly investigated for the case of Cooper pairs with spin-singlet symmetry. Here, we consider a spin-triplet superconductor with chiral pairing. By full self-consistent analysis of the spatial dependent superconducting order parameter, we determine the magnetic ground state of the superconducting spin valve. The study is performed by investigating the role of the orientation and strength of the magnetization in the ferromagnets including spin-valve asymmetries in the magnetic configurations. Due to the nonvanishing angular momentum of the spin-triplet Cooper pairs we demonstrate that the induced magnetic coupling has an anisotropic character and a structure that can favor collinear or noncollinear magnetic orientations, thus mimicking a magnetic interaction of the Heisenberg or Dzyaloshinskii-Moriya type, respectively. We investigate the role of the physical parameters controlling the character of the magnetic exchange: the amplitude of the magnetization in the ferromagnets and the length of the superconducting spacer in the spin valve. Our study demonstrates that spin-triplet superconductors can be employed to devise anisotropic magnetic exchange and to allow for transitions in the spin-valve state from a collinear to noncollinear magnetic configuration.  
\end{abstract}

\maketitle 

\section{Introduction}
The control and manipulation of electron spin in superconducting heterostructures are fundamental milestones in solid state physics, especially for the development of superconducting spintronics~\cite{Linder2015,Eschrig_2015}. The Cooper pairs wavefunction in superconductor can have different symmetry properties which are constrained by its fermionic nature. For instance, in superconductors with one type of carrier or electronic band at the Fermi level, apart from the conventional even-parity (e.g. $s$-wave) spin singlet channel, electron pairs with spin-1 angular momentum and orbital odd-parity (e.g. $p$-wave, $f$-wave, etc.) can be realized~\cite{Sigrist,TanakaPWave,ReadGreen,Ivanov,KitaevUnpaired,Maeno2012,SatoAndo}. In the latter case, the order parameter has a spin angular momentum, which thus makes the superconductor prone to a significant modification and non-trivial reconstruction upon the influence of magnetic fields or magnetic exchanges~\cite{Murakami,Dumi1,Dumi2,Wright,Mercaldo2016,Mercaldo2017,Mercaldo2018,Mercaldo2019} when integrated in magnetic heterostructures. 
Apart from the orbital effects related to the crystal wave-vector dependence of the superconducting order parameter, electronic bands with multiorbital character at the Fermi level can yield a pairing structure equipped with both spin and atomic-derived orbital moments. Independently of the sources of angular momentum contributing to the Cooper pairs structure, these types of superconductors are marked by distinct properties or responses to external perturbations that break either time or crystalline symmetries. Such properties include, among many, spin-sensitive Josephson transport~\cite{Sengupta,Yakovenko,Yakovenko1,Cuoco2008,Romano2010,Romano2010b,BrydonJunction,Annunziata2011,Gentile,Romano2013,Romano2017,Romano2016}, as well as superconduc\-ting spintro\-nics and spin polarized supercurrent~\cite{Linder2015,Romeo2013,Chung2018,Poniatowski2022,Chung2022}, magneto-electric effects~\cite{Romano2019,Mercaldo2019,Ojanen2012,Tkachov2017},  interface magnetism~\cite{Romano2013,Romano2017,Romano2016,Hu2021,Sengupta,Yakovenko,Yakovenko1}, and, more recently, superconducting orbitronics~\cite{mercaldo2023,Chirolli2022,Mercaldo2022,Mercaldo2020,fukaya20,fukaya22,Mercaldo2021_spectr}, with a potential impact extendable to devices for quantum computation~\cite{Cai2023,mercaldo2023}.

A remarkable manifestation of the interplay between magnetism and superconductivity is represented by the possibility of engineering the spin-exchange interaction among electrons through Cooper pairs. This physical scenario is typically encountered when dealing with superconducting spin-valve (SSV) consisting of ferromagnets (F) separated by a superconductor (S) as in FSF trilayers. 
The investigation of superconducting spin-valve has been originally proposed for the case of spin-singlet superconductors~\cite{Oh1997,Tagirov1999} with the aim to achieve a device that can act as a valve for the superconducting current flow by mainly exploiting the dependence of the superconducting critical temperature ($T_c$) \cite{Oh1997,Tagirov1999,Fominov2003,You2004,Fominov2010,Mironov2014,Gaifullin_2016} on the magnetic moment orientation in the ferromagnetic layers. In this framework, parallel and antiparallel magnetic orientations have been reported to yield positive~\cite{Gu2002,Moraru2006,Miao2008,Leksin2010,Zhu2010,Kehrle2012,Li2013,Zhu2017,Gu2015} or negative~\cite{Zhu2009,Rusanov2006,Steiner2006,Singh2007,Stamopoulos2007,Hwang2010,Flokstra2010,Kehrle2012} variation of $T_c$ as a consequence of the variety of involved mechanisms, the type of ferromagnets and the interface properties. Apart from the modification of the critical temperature, another core challenge refers to the control of the spin and charge transport in the FSF heterostructure~\cite{Yamashita2002,Takahashi2003,Krishtop2015,Gentile2022,Lu2022}. 

In this paper we focus on the character of the magnetic interaction mediated by the Cooper pairs in the superconductor of the spin-valve~\cite{Zhu2017,dibernardo2019}. The resulting coupling is substantially different from the super-exchange or double-exchange interactions encountered in magnetic materials. Indeed, those exchanges arise between electrons spin due to the effects of Coulomb interaction or from the Ruderman-Kittel-Kasuya-Yosida (RKKY) indirect exchange between localized spins 
mediated by itinerant electrons in metals.
A seminal work in the context of magnetic exchange mediated by Cooper pairs is due to de Gennes  ~\cite{degennes1966} who, inspired by the discovery of giant magnetoresistance, proposed a magnetic memory concept in which the superconducting transition temperature of a thin-film $s$-wave spin-singlet superconductor between ferromagnets is dependent on the relative magnetization alignment. 
{\textcolor{black}{The observation there is that a superconductor with size shorter than the coherence length can mediate the exchange between magnetic moments and that for spin singlet superconductor an antiferromagnetic-like coupling is obtained. This implies that, although the energy competition between the parallel and antiparallel configuration is not monotonous \cite{Halterman2005}, an antiparallel alignment tends to be energetically more favorable than a parallel one.} 
Along this line of investigation, the interaction between localized magnetic moments through dirty $s$-wave superconductors ~\cite{rkky1,rkky2,rkky3} has been worked out and shown to be marked by two main parts: one contribution is from the usual RKKY interaction and the second term with an exponential decay over the superconducting coherence length $\xi$. A similar scenario with an oscillatory term and one term favoring an anti-parallel configuration is also realized in SSV based on $d$-wave superconductors. 

In this context, the role of nodal excitations has been recently shown to be crucial in setting out both the sign and the length scale of the magnetic exchange in the superconducting spin valve~\cite{dibernardo2019} for a regime of a superconductor thickness that can be even larger than the coherence length. Indeed, in the case of $d$-wave pairing in heterostructures based on high temperature superconductors, 
one can achieve a ferromagnetic-to-antiferromagnetic transition by suitably varying the thickness of the superconductor or the strength of the proximity coupling between the magnetic and the superconducting layers~\cite{dibernardo2019}. For such magnetic heterostructures, the self-consistent determination of the superconducting order parameter as modified by the magnetic coupling is quite relevant. Indeed, the order parameter rearrangement can lead to counter-intuitive effects, as the fact of having a superconducting gap that is larger in the parallel configuration compared to the anti-parallel one, thus
increasing the superconducting condensation energy, even when the preferred ground state is marked by an 
anti-parallel magnetic alignment~\cite{Ghanbari2021,dibernardo2019}.

As discussed above, most of the studies on the influence of superconductivity on the magnetic state of the superconducting spin valve have been focusing on Cooper pairs with spin-singlet structure.
Thus, it turns out to be an open and general question to assess the impact on the magnetic state of superconductors whose Cooper pairs are instead marked by a non-vanishing angular momentum, for instance in the spin or in the orbital channel.   
In this paper, we face this problem and investigate which type of interaction between magnetic moments can be obtained in superconducting spin-valve based on spin-triplet superconductors focusing on the specific case of chiral pairing. 
It is indeed expected that due to the non-vanishing angular momentum of the spin-triplet Cooper pairs both in the spin and orbital part, as due to the chiral configuration, a direct coupling between the magnetic moments across the spin valve can be transferred through the condensate. The chiral spin-triplet case is particularly interesting in this framework because at the ferromagnet-superconductor interface the coupling between the spin-triplet {\bf d}-vector and the magnetic moment can be turned from parallel to perpendicular depending on the strength of the magnetization in the ferromagnet~\cite{Gentile,BrydonJunction}.
Here, we investigate the most favorable magnetic state that is realized in the SSV by taking into account the self-consistent rearrangement of the superconducting order parameter near the interfaces and in the inner side of the superconductor. 
We show that differently from the case of spin-singlet
superconductors, the magnetic ground state can be swapped from a configuration with parallel ($\phi=0$) or antiparallel ($\phi=\pi$) orientation of the magnetization to a state with perpendicular ($\phi=\pi/2$) relative orientation (see Fig.~\ref{f1}) of the magnetic moments. The modification of the magnetic state of the spin valve can be achieved by varying the strength of the magnetic exchange in the ferromagnets, the temperature or the size of the superconducting spacer. Our findings demonstrate that the spin-triplet Cooper pairs can mediate a Heisenberg or Dzyaloshinski-Moriya type of magnetic exchange. The study is performed for different size of the superconductor with respect to the coherence length and for various strengths of the magnetization including asymmetric magnetic configuration in the left and right leads of the superconducting spin valve.

The paper is organized as follows. In Sec. II, we describe the model and the methodology. 
In Sec. III, we discuss the behavior of the spin-triplet superconducting spin valve by considering the temperature dependence of the order parameter and the energetics regarding the competition of the magnetic states in the spin valve. The conclusions are reported in Sec. IV.

\begin{figure}[!t]
\includegraphics[width=0.99\columnwidth]{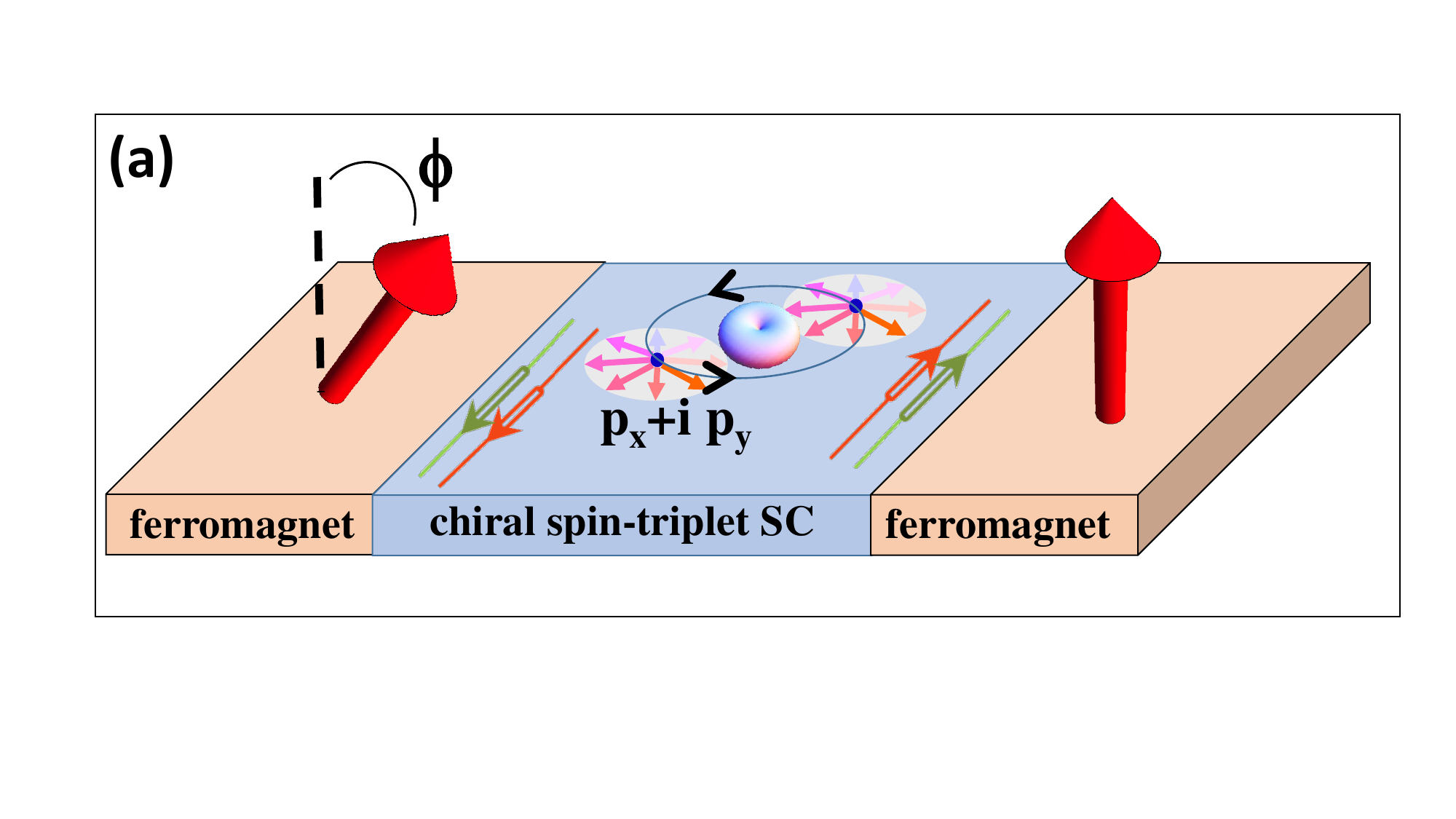}\vspace{-1.2cm}
\includegraphics[width=0.99\columnwidth]{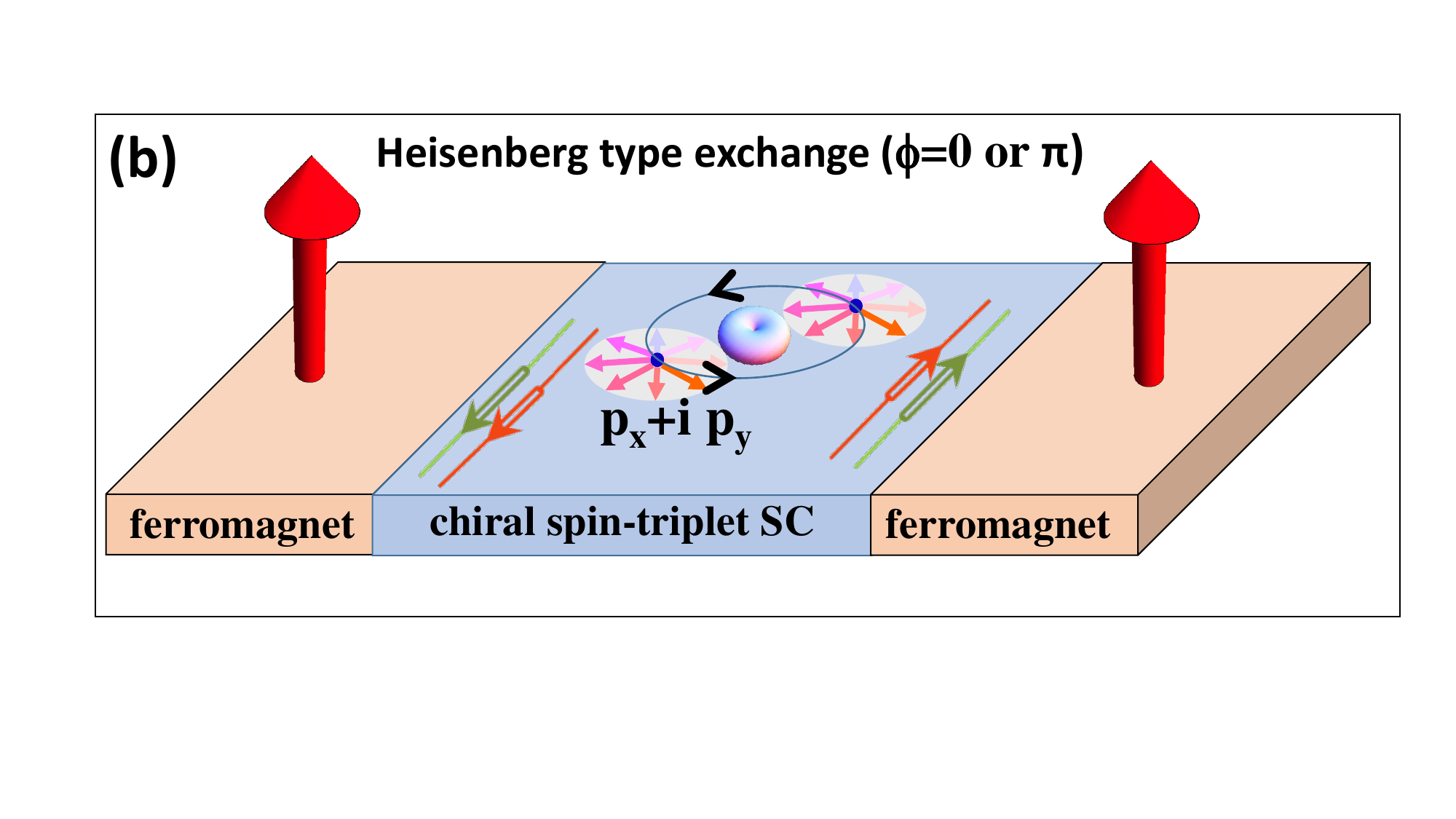}\vspace{-1.2cm}
\includegraphics[width=0.99\columnwidth]{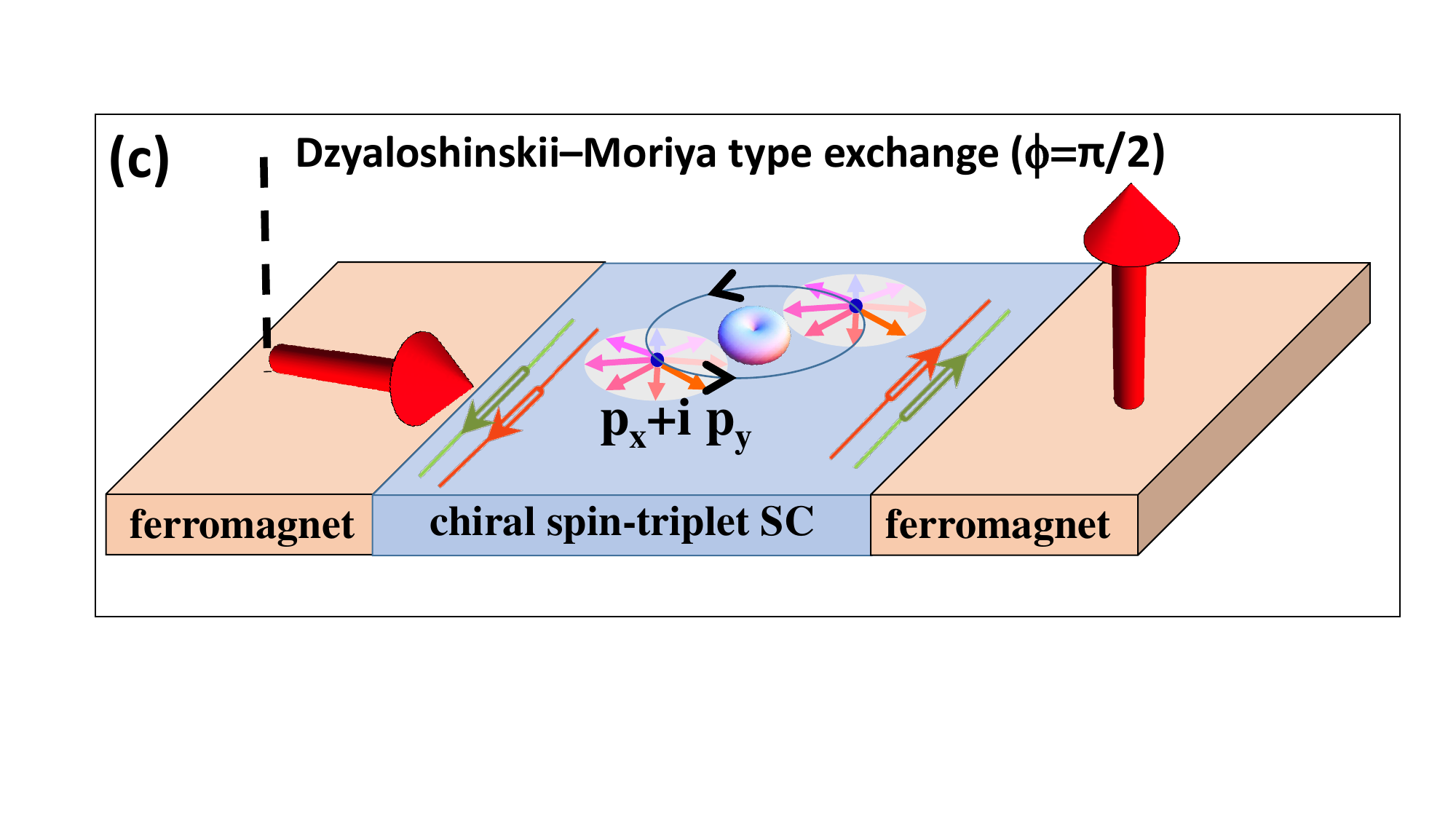}
\protect\caption{Schematic of the investigated spin-valve heterostructure with a chiral spin-triplet superconductor having $p_x+i p_y$ structure sandwiched between two ferromagnets. Magenta arrows in the middle layer denote parallel-spin Cooper pairs lying in the xy plane, while green and red arrows along the interfaces stand for spin-up and spin-down charge currents due to the chiral symmetry of the superconductor. 
The magnetizations $M_L$ and $M_R$ on the left and right side of the junction, respectively, can have a relative orientation marked by an angle $\phi$ with respect to the out-of-plane perpendicular direction as in the panel (a). (b) for $\phi=0$ ($\phi=\pi$) the magnetic moments $M_L$ and $M_R$ are ferromagnetically (antiferromagnetically) aligned, respectively. The effective magnetic exchange for parallel or anti-parallel alignment is of Heisenberg type. For the case of perpendicularly oriented magnetization (($\phi=\pi/2$) the exchange mimics the form of an effective Dzyaloshinski-Moriya interaction.}
\label{f1}
\end{figure} 

\section{Model and methodology}
We consider a planar heterostructure of size $L\times L$ (in units of the lattice constant $a$) extending in the $x$-$y$ plane, with two interfaces separating the chiral $p$-wave spin-triplet superconductor from the ferromagnetic leads. The ferromagnets are described by a conventional Stoner exchange whose strength is expressed by the amplitudes $h_{L}$ and $h_{R}$ on the left and right sides of the spin valve, respectively. 

We consider for simplicity the case of ferromagnetic layers having fixed equal width $d_F$, so that if we denote the lattice sites by $\mathbf{i}\equiv(i_x,i_y)$, with $i_x$ and $i_y$ being integers going from $-L/2$ to $L/2$, the two ferromagnet-superconductor interfaces are located at $i_x=\pm (L/2-d_F)$. The junction thus develops symmetrically with respect to the line ($i_x$=0,$\,i_y$). 

The Hamiltonian can be then expressed as

\begin{eqnarray} H&=& -t\sum_{\langle \mathbf{i}
,\mathbf{j} \rangle,\,\sigma} (c^{\dagger}_{\mathbf{i}\,\sigma}
c_{\mathbf{j}\,\sigma}+{\rm h.c.}) -\mu \sum_{\mathbf{i},\sigma}
n_{\mathbf{i}\sigma} \nonumber \\&& - \sum_{\langle \mathbf{i}
,\mathbf{j} \rangle \in \text{TSC}} V\left( n_{\mathbf{i}
\uparrow} n_{\mathbf{j}\downarrow}+n_{\mathbf{i}\downarrow}
n_{\mathbf{j} \uparrow} \right) - \sum_{\mathbf{i} \in \text{FM}}
\mathbf{h}\cdot\mathbf{s}_{\mathbf i}\;,
\end{eqnarray}
where $c_{\mathbf{i}\,\sigma}$ is the annihilation operator of an
electron with spin $\sigma$ at the site ${\mathbf{i}}$, $\langle
\mathbf{i},\mathbf{j}\rangle $ indicates nearest-neighbor sites,
$\mu$ is the chemical potential, and $\mathbf{s}_{\mathbf i} =
\sum_{s,s'}c^{\dagger}_{\mathbf{i}\,s}{\pmb{\sigma}}_{s,s'}c_{\mathbf{i}\,s'}$
is the spin density at site $\mathbf{i}$. The lattice is divided
into three regions: the two ferromagnetic (FM) subsystems for $|i_x|>L/2-d_F$, and the spin-triplet superconductor (TSC) 
subsystem for $|i_x|<L/2-d_F$. We assume that the hopping matrix element $t$ is the
same in the TSC and FM side of the system; relaxing this
assumption is not expected to qualitatively alter our results.
The charge transfer at the interface is given by $t_{{\text{
int}}}=\alpha t$ with $\alpha=1$, as we consider a regime of
\textcolor{black}{perfectly transparent interface}}. A variation of the FM-TSC interface transparency does not affect the magnetic behavior of the SSV. All energy scales are expressed in units of $t$. A nearest-neighbor attractive interaction $-V$ $(V>0)$ is
present only within the TSC side of the SSV. We choose the
electron density and pairing strength to get at mean-field
level a TSC state with the ${\bf d}$-vector parallel to the $z$-axis~\cite{Kuboki2001,Cuoco2008,Romano2010}.

Concerning the structure of the superconducting order parameter, in the $2\times2$ spin space the spin triplet configuration is commonly written in terms of the odd vectorial function $\bf d_{\bm k}$ in the form
${\Delta}_{\bm k}=i\left(\bf d_{\bm k}\cdot\bm\sigma\right)\sigma_y$ ~\cite{Sigrist},
where $\sigma_i \; (i=x,y,z)$ are the Pauli matrices. In the following, we consider a triplet vector $\bf d_{\bm k}$ 
of chiral type, ${\bf d_{\bm k}} \equiv (0, 0, \sin(k_x)+i \sin(k_y))$, i.e. $\bf d_{\bm k}$ is oriented along the $z$-direction.
{\textcolor{black}{To introduce the superconducting order parameter for the spin-valve geometry, we recall that one decouples the interaction term  by introducing the pairing amplitude $\Delta_{{\bf{i,j}}}=\langle c_{{\bf{i}}\uparrow} c_{{\bf{j}}\downarrow} \rangle$ on a given bond, involving only nearest-neighbor sites ${\bf{i}}$ and ${\bf{j}}$ on the lattice~\cite{Kuboki2001,Cuoco2008,Romano2010}, so that 
\begin{equation*}
V_{{\bf{i,j}}} n_{{\bf{i}}\uparrow} n_{{\bf{j}}\downarrow} \cong V_{{\bf{i,j}}} \left[ \Delta_{{\bf{i,j}}}
c^{\dagger}_{{\bf{j}}\downarrow} c^{\dagger}_{{\bf{i}}\uparrow}+ \Delta^{*}_{{\bf{i,j}}} c^{\dagger}_{{\bf{i}}\uparrow} c^{\dagger}_{{\bf{i}}\downarrow} - |\Delta_{{\bf{i,j}}}|^2 \right]  
\end{equation*}
where the average in the definition of $\Delta_{{\bf{i,j}}}$ indicates the finite temperature expectation amplitude. After the decoupling the Hamiltonian has a bilinear form, $H_{MF}$, which can be diagonalized by means of standard numerical routines. As mentioned above, for our analysis we assume that $V_{{\bf{i,j}}}$ is  not vanishing only for nearest-neighbor sites and that the pairing strength is the same for the $x$ and $y$ directions, i.e. $V_{{\bf{i,i+a_x}}}=V_x=V$ and $V_{{\bf{i,i+a_y}}}=V_y=V$, ${\bf{a_x (a_y)}}$ being the unit vectors connecting the nearest-neighbor sites along the direction perpendicular (parallel) to the interface. To describe the spin-triplet superconductor, the pairing amplitudes on each bond can be combined to yield the spin triplet superconducting order parameter in the $S_z=0$ channel ($d_z$ component). Indeed, we have that the spin-triplet order parameter $\Delta^{T}$ on the bond is constructed by the antisymmetric combination $\Delta^{T}_{\bf{i,j}}=\frac{1}{2}(\Delta_{{\bf{i,j}}}-\Delta_{{\bf{j,i}}})$. Hence, we can define the superconducting order parameters, at a given site ${\bf i}$, with p-wave symmetry along the $x$ and $y$ directions as
\begin{equation*}
p_x(p_y)({\bf{i}})=\frac{1}{2}(\Delta^{T}_{{\bf{i,i+a_x (a_y)}}}-\Delta^{T}_{{\bf{i,i-a_x (a_y)}}}) \,. 
\end{equation*}
Since we have translational symmetry along the $y$ direction parallel to the interface, one can perform a Fourier transform and introduce the momentum $k_y$ to evaluate the average of the lateral dimension of the heterostructure. Then, the dependence on the site position will be explicit only for the $x$ coordinate along the direction perpendicular to the interface. The pairing amplitude is evaluated by an iterative approach until the desired accuracy is reached (see Appendix A for details).}}


The FM subsystems are modelled by the exchange field ${\bf h}$ whose orientation determines the magnetization direction. 
We consider various configuration for ${\bf h}$ in the left and right side of the SSV. In particular, we introduce the polar angle $\phi$ to set the orientation of the magnetization with respect to the out-of-plane direction $z$ in the left side ferromagnet. We assume a biaxial anisotropy for the isolated ferromagnet by considering that the ferromagnetic configurations with magnetic moments along the $z$-axis or a given direction within the $x-y$ plane, e.g. $x$-axis, are degenerate in energy. 
Then, in order to assess the optimal magnetic state of the SSV we compare the free energy for five distinct magnetic states. This is done by considering various magnetic configurations with parallel, antiparallel or perpendicular relative orientations of the magnetization in the left and right side of the SSV, both along the $z$- and $x$-axis. For convenience, we use for them the notation $(\sigma,\sigma')$ with $\sigma,\sigma'=\uparrow,\downarrow \; (\rightarrow,\leftarrow)$ for the $z$-oriented ($x$-oriented)  configurations in the two ferromagnetic layers.


\begin{figure}[!t]
\centering
\includegraphics[width=0.99\columnwidth]{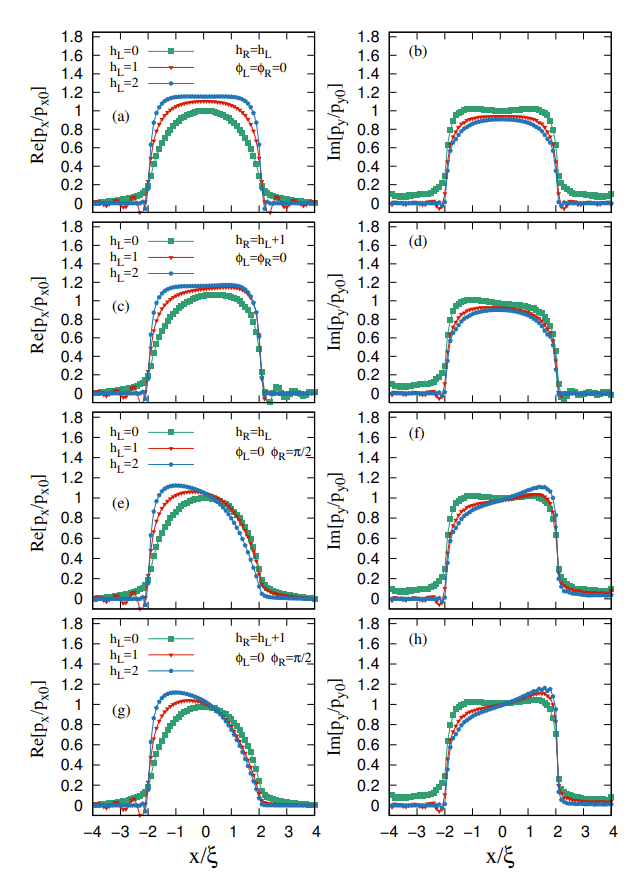}
\protect\caption{Zero-temperature superconducting order parameter for several choices of the exchange fields in the left (L) and in the right (R) layer, respectively: equal magnitude and same direction, parallel to the $\bf d$-vector (panels a) and b)); different magnitude and same direction, parallel to the $\bf d$-vector (panels c) and d)); equal magnitude and directions parallel and perpendicular to the $\bf d$-vector, respectively (panels e) and f)); different magnitude and directions parallel and perpendicular to the $\bf d$-vector, respectively (panels g) and h)). The order parameters are normalized to the value that they assume in the middle of the superconducting layer, in the case of normal-superconductor-normal (NSN) junction. $\xi=10 a$ is the zero-temperature coherence length, $a$ being the lattice constant.}      
\label{f2}
\end{figure}

\section{Results}

In this Section we present the spatial and thermal profile of the spin-triplet superconducting order parameter for different spin-valve magnetic configurations and thickness of the superconducting layers. Moreover, we perform a thorough analysis of the energetics to assess the most favorable magnetic state of the spin-valve.


\subsection{Proximity effects and superconducting order parameters}

We start by considering the proximity effect in the ferromagnet-triplet superconductor-ferromagnet spin-valve by focusing on the corresponding thermal and spatial evolution of the spin-triplet superconducting order parameter (Fig.~\ref{f2}),  
as obtained by means of spatial dependent self-consistent approach. As expected, the spatial profile of the order parameters is significantly modified by the strength of the magnetic moment in the F side of the junction ~\cite{Gentile}. The spatial profile  also depends on whether one considers $p_x$ or $p_y$ components of the chiral spin-triplet order parameter. Here, we take two representative configurations with the magnetization that is either aligned ($\phi=0$) or perpendicular ($\phi=\pi/2$) to the $\bf d$-vector and we consider both symmetric and asymmetric amplitudes of the magnetization in the ferromagnets.
Starting from the zero magnetic exchange configuration (i.e. $h_R=h_L=0)$) with the magnetization aligned parallel to the $\bf d$-vector (Fig.~\ref{f2} (a,b)), we have that the $p_x$ and $p_y$ components exhibit a standard spatial profile with a monotonous decaying amplitude within the ferromagnetic region, as expected in conventional normal-superconductor heterostructures. It is worth noticing that the amplitude of the $p_x$ component inside the superconducting region gets suppressed at the interface because the order parameter is odd in momentum and in the Andreev reflection at the FT interface such sign change yields a reduction of the pairing amplitude. On the other hand, the $p_y$ component is not affected by the interface. We recall that for the examined interface the momentum $k_y$ is conserved.

In the presence of a nonvanishing magnetic moment in the ferromagnet, the spatial dependence of the order parameter is characterized by some distinctive features in the ferromagnet and in the superconductor region of the spin-valve, respectively.
Regarding the ferromagnetic side, one finds that the order parameter has a profile that is spatially oscillating for the case of the magnetization being collinear to the $\bf d$-vector (Fig.~\ref{f2}) along the out-of-plane direction. This behavior arises from the fact that the exchange field is pair breaking for the spin-triplet configuration with spins lying in the $xy$ plane. The characteristic length scale of the oscillation of the order parameter becomes shorter with the increase of the magnetic exchange strength, similarly to what is obtained for the case of a ferromagnet-superconductor heterostructure with spin-singlet Cooper pairs. Such behavior is observed for both the $p_x$ and the $p_y$ components of the order parameter. 
Instead, for a magnetization that is perpendicular to the $\bf d$-vector, the pairing amplitude has a monotonous decay moving away from the FT interface toward the inside of the ferromagnet (Fig.~\ref{f2}). Additionally, as expected, the pairing amplitude gets suppressed with the increase of the magnetization in the ferromagnet.  
Considering now the behavior of the order parameter in the superconducting region, we see that the proximity to the ferromagnet has a peculiar impact on its $p_x$ and $p_y$ components. In fact, the amplitude of the $p_x$ ($p_y$) component increases (decreases) at the interface when the magnetization is oriented parallel to the $\bf d$-vector (Fig.~\ref{f2} (e),(f)). 
On the other hand, the behavior is just opposite in the case of the magnetization being in the plane, perpendicular to the $\bf d$-vector (Fig.~\ref{f2} (g),(h)). The simulation of a spin-valve configuration with an out-of-plane and in-plane orientation of the magnetization for the left and right side, respectively, allows to track the behavior of the proximity effect at each interface. We point out that an asymmetric amplitude of the exchange field in the two sides of the spin-valve does not affect qualitatively the behaviour of the pairing amplitude both in the superconducting and ferromagnetic regions.
This result is demonstrated in Figs. \ref{f2}~(c), (d), (g), (h), where we have determined the behavior of the order parameter for several values of the magnetic exchanges on the two sides of the spin-valve assuming a difference $h_R-h_L=1$ (in units of the nearest-neighbor hopping amplitude). 

{\textcolor{black}{We would like to point out that the different behavior of the $p_x$ and $p_y$ components arises from the spin dependent Cooper pair reflection at the interface~\cite{Brydon2009,Brydon2011,Gentile}.
Indeed, assuming that $\phi$ is the angle between the $d$-vector and the magnetization $M$ in the ferromagnet, in the scattering process an incident Cooper pair with spin $\sigma$ mutually perpendicular to the $d$-vector and to $M$ acquires the spin-dependent $\eta_s$ and orbital-dependent $\eta_o$ phase shifts given by $\eta_s=\pi -2 \sigma \phi$ and $\eta_o=\Delta \theta$, respectively. Then, $\Delta \theta$ is the phase change of the superconducting gap upon specular reflection and, due to the orbital dependent shift, it is $\Delta \theta=\pi (0)$ for the $p_x (p_y)$ pairing state. Since the gap is suppressed at interfaces where reflected Cooper pairs acquire a nontrivial phase shift \cite{Buchholtz1981}, at the TSC-FM interface the gap is maximized by choosing the angle $\alpha$ such that in the scattering the Cooper pairs acquire a trivial phase shift that is a multiple of $2 \pi$. Because of the different orbital phase shifts $\Delta \theta$, the behavior of the superconducting order parameter for $p_x$ and $p_y$ is inequivalent.}}


\begin{figure}[!t]
\centering
\includegraphics[
width=0.9\columnwidth]{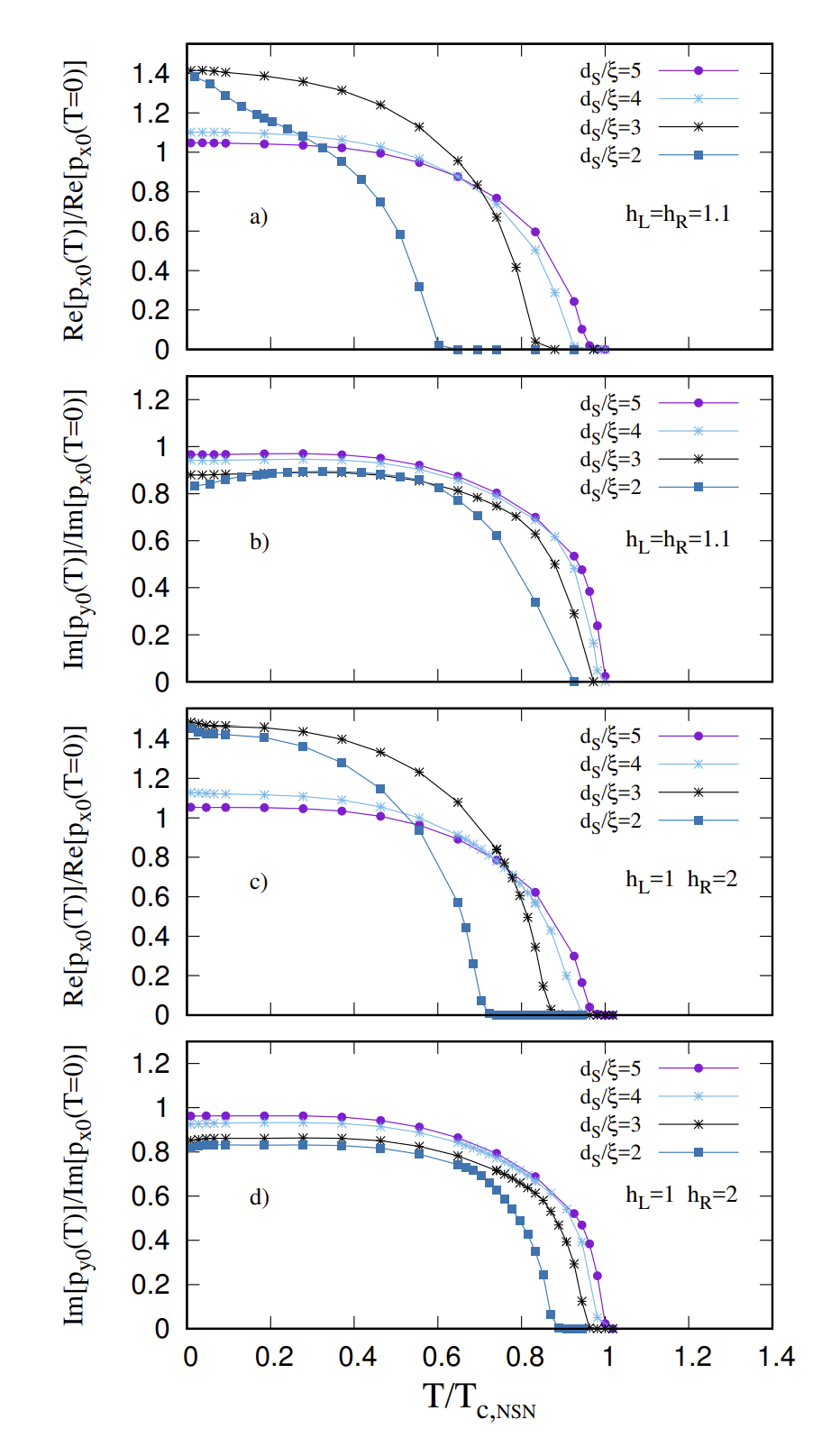}
\protect\caption{Temperature dependence of the magnitude of the triplet order parameter at the center of the superconducting layer for different thicknesses $d_S/\xi$ of the latter, $\xi$ being the zero-temperature superconductor coherence length: panels a) and b) show $p_x$-real and $p_y$-imaginary parts in the case of equal exchange field in the two ferromagnetic layers, respectively, whereas panels c) and d) show the same quantities in the case of different magnetic fields. $T_{c,{\rm NSN}}$ is the critical temperature in the NSN case for a thickness $d_S/\xi=4$ and the order parameters are normalized to their zero-temperature values in the NSN case. The exchange fields in the ferromagnetic layers are both aligned to the {\bf d}-vector thus realizing a z-oriented parallel spin valve configuration (e.g. $\uparrow,\uparrow$).}  
\label{f3}
\end{figure} 
\begin{figure}[!hbt]
\includegraphics[width=1.\columnwidth]{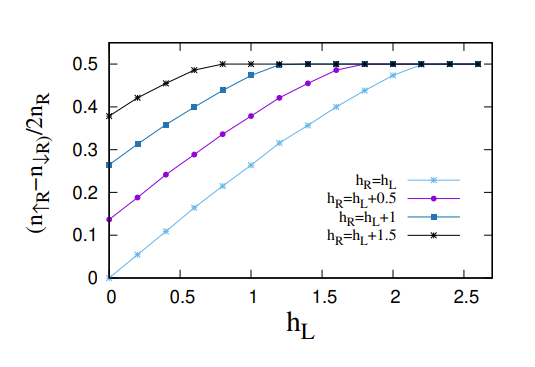}
\protect\caption{Magnetization evolution in the right ferromagnetic layer for $T/T_c=0.16$, $d_S/\xi=4$ and different choices of the magnetic exchanges $h_L$ and $h_R$.}
\label{f4}
\end{figure} 

Then, we analyze the temperature dependence of the order parameter for different magnetic configurations of the spin-valve and for various thickness of the superconducting spacer between the ferromagnets (Fig.~\ref{f3}). Here, for the parameters that set out the strength of the superconductor, we introduce the coherence length $\xi$, given by the variation of the order parameter when the superconductor is interfaced with the vacuum at zero temperature (in our case $\xi \simeq 10 \,a$). Then, we consider superconductors with different lateral thickness $d_s$ along the direction perpendicular to the interface ($d_s/\xi=2,3,4,5$). This analysis permits to investigate the way the ferromagnets gets coupled ranging from a thin superconductor regime in which the interfaces are close together and strongly connected to each other to a regime in which the electronic processes at the interface are transmitted through a region of the superconductor that behaves differently from the interface. Since we are dealing with a chiral spin-triplet superconductor, the breaking of the $C_4$ rotational symmetry in the spin-valve has a significant impact on the temperature dependence of the $p_x$ and $p_y$ components. Here, we consider a representative symmetric (i.e. $h_L=h_R=1)$ and asymmetric (i.e. $h_L=1,h_R=2)$ ferromagnetic configuration to evaluate the impact of the magnetic proximity in the spin-valve on the chiral order parameter. 
As expected, since the $p_x$ component is related to the direction of the superconductor's thickness reduction, its amplitude has a nontrivial evolution when changing $d_s$ due to the presence of the lateral ferromagnets. 
Indeed, the zero-temperature strength of the order parameter increases by reducing the thickness down to $d_s=3 \xi$ before there is a breakdown at a critical size $d_s=2 \xi$ (see Fig.~\ref{f3} (a)) until it vanishes. Contrary to the nonmonotonous trend of the zero temperature amplitude of the $p_x$ component, the critical temperature, below which the corresponding pairing gets switched on, exhibits a monotonous reduction of about 40$\%$ (see Fig.~\ref{f3}(a)). Hence, the resulting behavior of the order parameter does not follow the standard BCS behavior where the pairing amplitude scales with the critical temperature. {\textcolor{black}{Here, the superconducting transition temperature is scaled to the value $T_{\text{c,NSN}}$ associated with the normal-superconductor-normal (NSN) junction for $d_s=4 \xi$ which is the reference value we used for the unperturbed chiral superconductor without the coupling to the ferromagnets}}. 

\begin{figure}[htb]
\includegraphics[width=1.0\columnwidth]{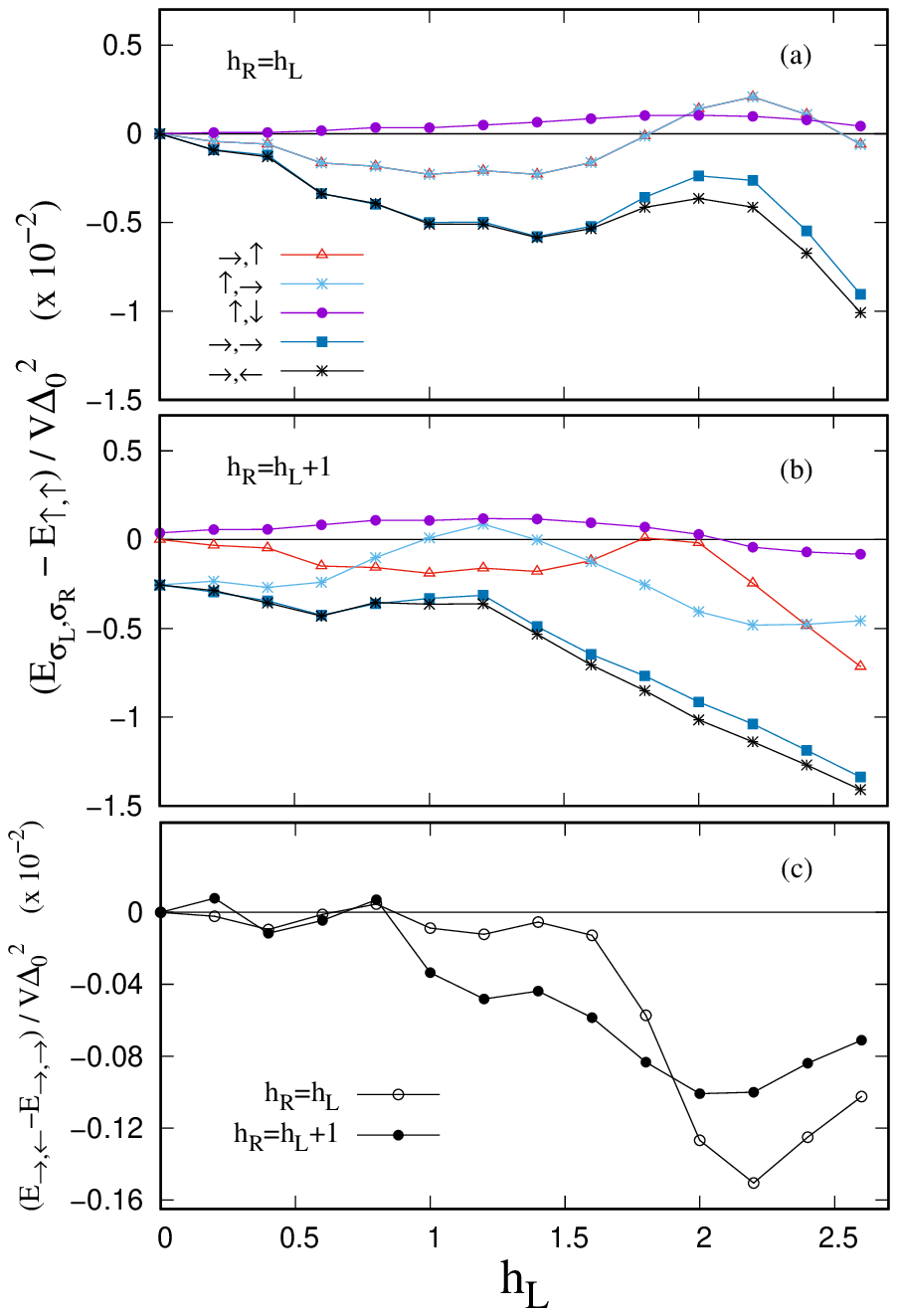}
\protect\caption{Free energy evolution of the spin valve magnetic states at $T/T_c=0.16$ and $d_S/\xi=3$ for symmetrical (panel a)) and asymmetrical (panel b)) amplitudes of the exchange fields in the two ferromagnetic layers. Energies of the different magnetic configurations are measured with respect to the case of a magnetic state with out-of-plane magnetizations, both parallel to the $\bf d$-vector. Panel c) reports the difference between the energies of the parallel and anti-parallel in-plane configurations.} 
\label{f5}
\end{figure} 

The temperature dependence of the $p_x$ order parameter is quite peculiar for a system size $d_s=2 \xi$. It exhibits a reentrant profile, setting in below $T \sim 0.6~T_{\text{c,NSN}}$ and then  vanishing again as zero temperature is approached. As expected, in this case of thin superconductor the order parameter is particularly sensitive to the boundary conditions. Indeed, for the case of a spin-valve with asymmetric amplitude of the magnetization (Fig.~\ref{f3}(c)) the reentrant behavior is not observed. Such outcome is related to the fact that the $p_x$ component is enhanced close to the interface by the increase of the magnetization.  

Unlike the $p_x$  order parameter, the component $p_y$ parallel to the interface is not significantly affected by the presence of ferromagnets. The critical temperature decreases as a function of the width of the superconductor for both the symmetrical and asymmetric configuration of the order parameter (Fig.~\ref{f3}(b),(d)). We observe that the zero temperature amplitude exhibits, instead, a nonmonotonous trend as a function of $d_s$, but only for the symmetrical spin-valve configuration. Again, the anomaly occurs for a thickness that is one or two times the value of the coherence length $\xi$ (Fig.~\ref{f3}(b)).
We have seen that the two components of the chiral order parameter have an activation at different temperatures depending on the width and on the orientation of the magnetic moments at the two sides of the spin-valve. Thus, the results demonstrate that it is possible to observe a series of transitions from a non-chiral to a chiral phase as a function of temperature by appropriately modifying the size of the superconductor or the magnetic configuration.

Before proceeding further on the energetics of the spin-valve, it is useful to describe how the magnetic exchange strength is related to the amplitude of the magnetization in the ferromagnet. This helps to identify the regimes of weak and strong strength of the ferromagnetic correlations. Additionally, we can set the amplitude of the exchange field above which the ferromagnet behaves as a half-metal, with only one spin orientation that is occupied at the Fermi level while the spin minority band is empty. This regime is relevant for singling out the optimal magnetic orientation to be parallel or perpendicular to the the $\bf d$-vector in the chiral spin-triplet superconductor. In Fig.~\ref{f4} we determine how the magnetization in the right ferromagnetic layer evolves from the unpolarized configuration in the left one ($h_L=0$) to the the half-metallic regime for different asymmetric amplitudes of the magnetic exchange. We find that the magnetization has a linear dependence on $h_L$, with a slight deviation from linearity near the half-metal regime, reached in the symmetric configuration for $h_L \sim 2$ (in units of the hopping amplitude). This threshold value of $h_L$ is, as expected, gradually reduced by the shift of the magnetization induced by the offset in the magnetic exchange. We point out that even though the results presented in Fig.~\ref{f4} refer to a representative value of temperature and thickness, different choices for them do not alter the trend of the magnetization discussed above.


\subsection{Energy competition: collinear vs noncollinear magnetic states}

Let us now consider the energetics of the spin-valve. By self-consistently determining the profile of the superconducting order parameter one can evaluate the energy for any given magnetic configuration as a function of the relative orientation of the magnetizations in the two ferromagnets. We first address the behavior of the spin-valve in terms of the exchange field strength taking a representative temperature and evaluating the ground state for different values of the thickness of the superconducting spacer. We point out that the minimum of the free energy generally corresponds to configurations with the magnetization in the left and right sides of the spin-valve that are aligned or anti-aligned along the out-of-plane or in-plane directions. Additionally, a local minimum of the free energy can also occur when the relative angle among the magnetizations is $\phi=\pi/2$ (Fig.~\ref{f1}) keeping the preferred orientations along the out-of-plane ($z$) or in-plane ($xy$) directions. To schematically indicate these magnetic configurations we introduce the following notation: $\uparrow$ and  $\downarrow$ stand for upward and downward orientation of the out-of-plane magnetization, respectively, while $\rightarrow$ and $\leftarrow$ label configurations with opposite oriented magnetization lying in the $xy$ plane. For the symmetry of the examined problem there is an angular degeneracy for all the in-plane orientations of the magnetization. Then, the spin-orbit coupling can lift the degeneracy by selecting only one preferential angle for the in-plane magnetic orientation. Nonetheless, we do expect that the qualitative outcome of the analysis is not altered by the change in the magnetic anisotropy.

\begin{figure}[t]
\includegraphics[width=1.\columnwidth]{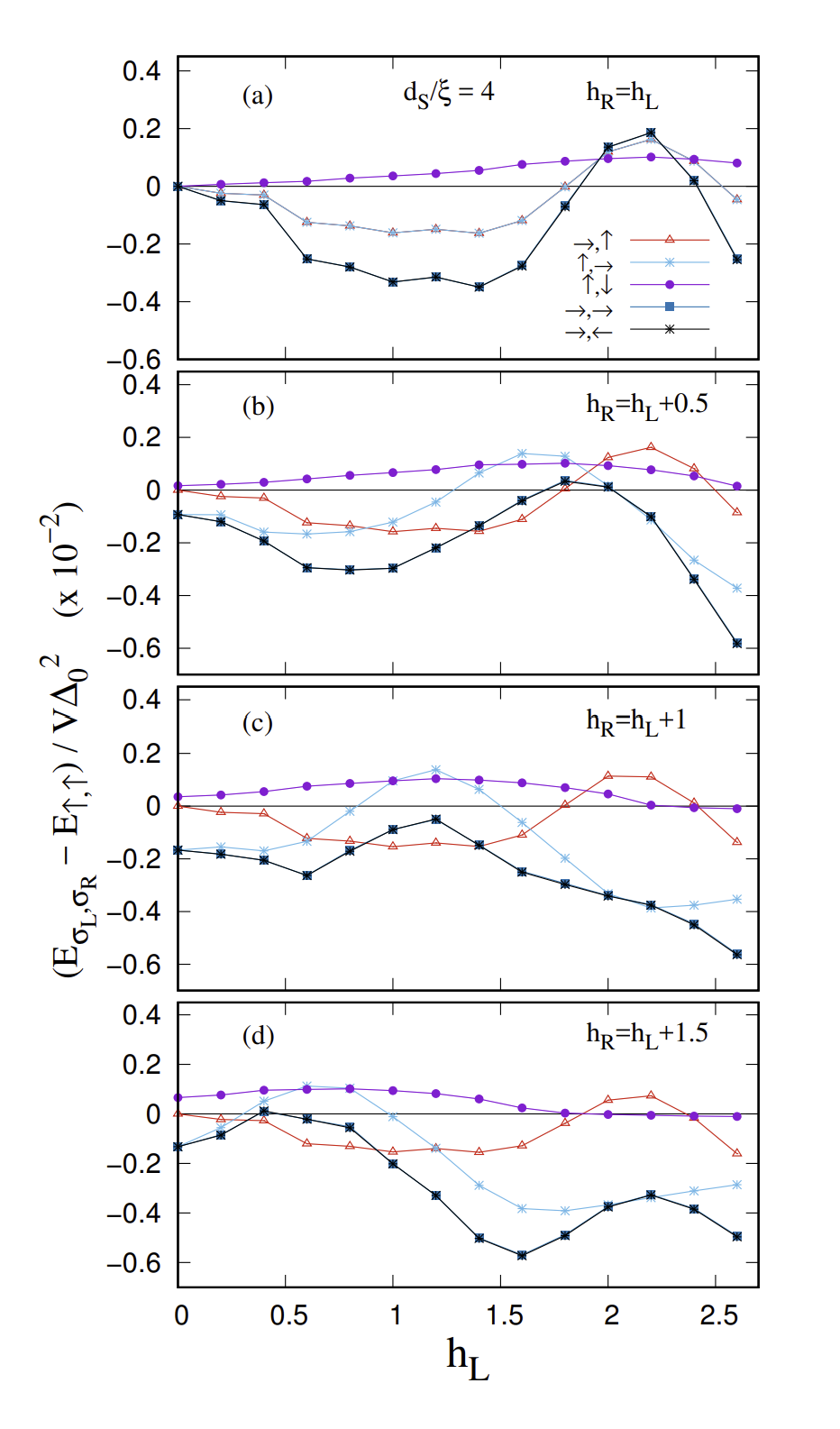}
\protect\caption{Same as in Fig.~\ref{f5} for $T/T_c=0.16$, $d_S/\xi=4$ and different choices of the exchange field in the two ferromagnetic layers. We observe that the main competition for the ground state of the spin valve is between the antiferromagnetic planar configuration ($\rightarrow,\leftarrow$) and the perpendicularly oriented magnetic state ($\uparrow,\leftarrow$).}
\label{f6}
\end{figure} 

We start from the case of a superconducting spacer with thickness $d_s= 3 \xi$ (Fig.~\ref{f5}) by focusing on the regime of low temperature ($T/T_c  \leq 0.2$). In this case there are two possible physical situations which are related to the strength of the magnetization. For the case of weak ferromagnetism ($h_L \leq 0.6)$ and a magnetization that does not exceed 0.25 $\mu_B$, the preferential orientation of the magnetic moment is in the plane of the spin valve, with 
no significant energy separation among the ferromagnetic ($\rightarrow,\rightarrow$) and the antiferromagnetic ($\rightarrow,\leftarrow$) configurations. 
This behavior does not depend on whether the spin valve is in a configuration with symmetric or asymmetric amplitudes for the magnetic moments in the left and right ferromagnets. 
When higher values of the fields are considered, the energy difference between the two in-plane magnetic configurations becomes more significant, with the antiparallel one being the lowest.     

\begin{figure}[!t]
\includegraphics[width=0.95\columnwidth]{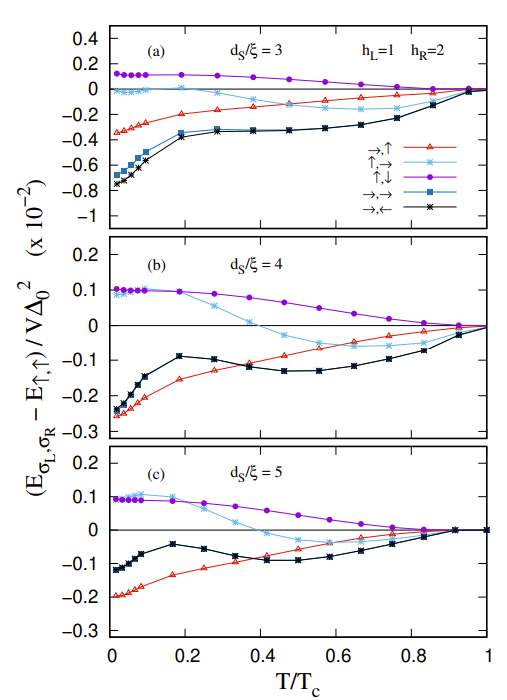}
\protect\caption{Temperature evolution of the ground-state energy for different magnetic configurations and different thicknesses $d_S/\xi$ of the superconductiong layer. Energies are measured with respect to the one of the case of exchange fields both parallel to the ${\bf d}$-vector, and are normalized to $V\Delta_0^2$, $\Delta_0$ being the zero-temperature order parameter in the NSN (normal-superconductor-normal) junction. Vertical (horizontal) arrows in the legend denote exchange fields in the two ferromagnetic layers perpendicular (parallel) to the planar junction.}
\label{f7}
\end{figure} 

The ground state of the spin-valve is instead significantly modified by the increase of the superconducting thickness. In particular, the asymmetric configurations display a rich behavior when the strength of the exchange field is varied (Fig.~\ref{f6}).
To exemplify the behavior of the spin valve in this thickness regime, let us start by considering the symmetric exchange configuration (i.e. $h_L=h_R$). We first observe that, although the lowest energy configuration for most of the exchange amplitudes is the one with the magnetic moments lying in the $xy$ plane and coupled antiferromagnetically ($\rightarrow,\leftarrow$), there is a non trivial window of values near $h_L\sim 2$ with the ground state having out-of-plane moments that are antiparallel ($\uparrow,\downarrow$).
On the other hand, for asymmetric amplitudes of the magnetizations on the left and right side of the spin-valve other transitions can be observed.
Indeed, one can induce a transition from an antiferromagnetic state with in-plane magnetic moments, i.e. $(\rightarrow,\leftarrow)$, to a noncollinear configuration with the magnetizations that are perpendicularly oriented as ($\uparrow,\leftarrow$). This type of ground state for the spin valve is obtained only in a finite range of values of $h_L$ assuming that a finite difference between $h_L$ and $h_R$ is set, implying an asymmetric exchange between the left and right ferromagnets in the spin-valve, as shown in Fig.~\ref{f6}. The possibility of achieving a noncollinear configuration is substantially due to the fact that at the ferromagnet-superconductor interface the chiral superconductor can energetically favor either a magnetization that is parallel or perpendicular to the $\bf d$-vector. In particular, the parallel configuration (${\bf M} \parallel {\bf d}$) is achieved when the magnetization amplitude approaches the half-metallic regime. For this reason, if the magnetization is large there will be an energy gain for ${\bf M} \parallel {\bf d}$ while for the weak ferromagnetic configuration the state ${\bf M} \perp {\bf d}$ is favored. Hence, for dominant interface effects one can realize a spin valve with noncollinear magnetization by suitably selecting the amplitude of the magnetic exchange. Such physical case occurs for a sufficiently large thickness of the superconductor.

Let us now consider the role of temperature. As we have shown in Fig.~\ref{f3} the spatial profile of the order parameter is significantly dependent on the thickness of the superconducting spacer and on the strength and orientation of the magnetization in the ferromagnets. 
For this reason, it is particularly interesting to investigate how the energy of the various magnetic configurations evolve with temperature in order to identify the most favorable magnetic state and the hierarchy in energy among the collinear and noncollinear states.
In Fig.~\ref{f7} we take a representative asymmetrical configuration of the spin valve (i.e. $h_L=1$ and $h_R=2$) and consider the evolution of the magnetic states from zero temperature to the transition temperature. For clarity the free energy of each spin valve state is evaluated with respect to that with parallel out-of-plane magnetic moments ($\uparrow,\uparrow)$. 
We start by observing that the antiferromagnetic spin-valve state with out-of-plane magnetization, ($\uparrow,\downarrow)$, is substantially the highest in energy as compared to the configurations with planar magnetization or with perpendicularly oriented magnetic moments. This is because for a magnetization parallel to the $\bf d$-vector pair breaking mechanisms tend to dominate and thus they suppress the energy gain arising from the superconducting order parameter.
Regarding the temperature dependence of the spin-valve configuration, we find that it is possible to achieve a thermally driven magnetic transition but only for thickness that are greater than $d_s/\xi=3$.
For thin spin-valve ($d_s/\xi=3)$ the magnetic ground-state is given by the planar antiferromagnetic configuration $(\rightarrow,\leftarrow)$. While the latter is clearly separated in energy from the noncollinear configurations or the states with out-of-plane magnetic moments, it is strongly competing with the ferromagnetic planar state $(\rightarrow,\rightarrow)$ (Fig.~\ref{f7} (a)). They are indeed about degenerate in energy for temperatures ranging from $T \sim 0.5 T_c$ to the transition temperature.
This behavior indicates a nontrivial role of the electronic excitations in favoring a ferromagnetic coupling among the ferromagnets. 
Regarding the case of a thicker superconducting spacer, we have that for $d_s/\xi=4$ and $d_s/\xi=5$ the spin-valve undergoes a transition at $T \sim 0.4 T_c$ from a perpendicularly oriented magnetic state, $(\rightarrow,\uparrow)$, to an in plane collinear state with anti-aligned magnetic moments $(\rightarrow,\leftarrow)$ (Fig.~\ref{f7} (b),(c)). 
In this range of values of $d_s/\xi$, the local interface coupling starts playing a role in setting out the magnetic ground state of the spin-valve. Indeed, while for shorter superconducting spacer the direct magnetic exchange mediated by the superconductor favors the antiferromagnetic planar state at low temperatures, the increase of the superconducting thickness weakens this interaction and the interface coupling between the $\bf d$-vector and the magnetization becomes more relevant.
The transition can be then qualitatively accounted for by the fact that the coherence length of the superconductor depends on temperature and it grows approaching the transition into the normal state. Then, at higher temperatures the direct exchange among the magnetic moments dominates over the interface coupling between the $\bf d$-vector and the magnetization, thus stabilizing the planar configuration again. We point out that the energy separation between the planar ferromagnetic and antiferromagnetic states gets reduced by increasing the thickness of the superconducting spacer.

\begin{figure}[!t]
\includegraphics[width=0.95\columnwidth]{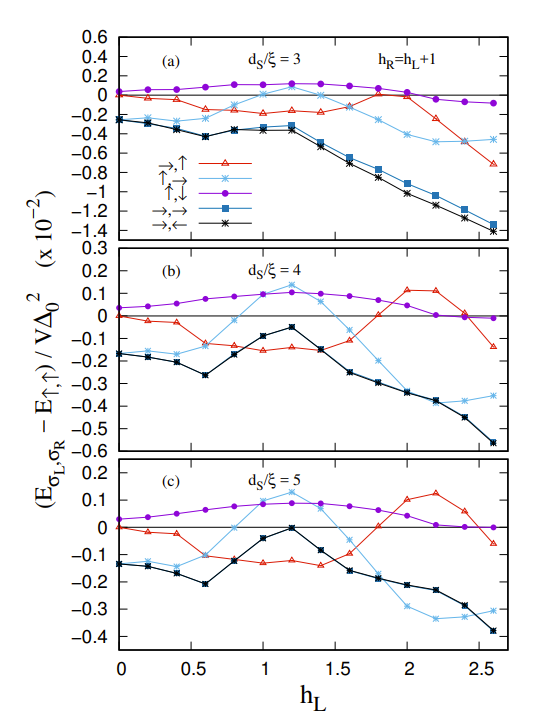}
\protect\caption{Same as in Fig.~\ref{f5} for different values $d_S/\xi$ of the thickness of the superconducting layer at a representative value of the temperature $T/T_c=0.16$.  }
\label{f8}
\end{figure}

To further assess the interplay among the collinear and noncollinear configurations at low temperatures we track the competition of the magnetic ground states in terms of thickness and exchange field asymmetry in the ferromagnets (Fig.~\ref{f8}). Here, we first observe that for thin superconducting spacer ($d_s/\xi=3$) the antiferromagnetic planar configuration ($\rightarrow,\leftarrow$) is the ground state for any amplitude of the magnetic exchange (Fig.~\ref{f8}(a), same as Fig.~\ref{f5}(b)). This in-plane antiferromagnetic configuration is close in energy to the ferromagnetic one with in-plane parallel magnetic moments and the splitting becomes sizable with the increase of the amplitude of the exchange field. It is interesting to notice that, in contrast to the behavior of the magnetic planar state, for the out-of-plane collinear configurations the antiparallel state ($\uparrow,\downarrow)$ is higher in energy compared to the parallel state $(\uparrow,\uparrow)$ (or equivalently $(\downarrow,\downarrow)$). We argue that this finding is due to the fact that equal spin Cooper pairs associated to the $d_x$ and $d_y$ component of the $\bf d$-vector can be induced in the spin-triplet superconductor by inverse proximity~\cite{Maistrenko2021} and the free energy is optimized by having only one dominant component. The induced pair correlations may indeed have an impact on the spectrum of the superconductor both in the gap and above the gap and thus determine the most favorable magnetic configuration.   
The increase of the superconducting layer thickness introduces the possibility of having a transition from the collinear to noncollinear configuration in a limited range of exchange fields corresponding to a half-metallic magnetization for one of the ferromagnet (Fig.~\ref{f8}(b),(c)). The noncollinear $(\rightarrow,\uparrow)$ configuration arises from the interface coupling between the magnetic moments and the $\bf d$-vector that on the right ferromagnet in the spin-valve, due to the large magnetization (see Fig.~\ref{f5}), is energetically more favorable. It is interesting to notice that when $h_L$ is large enough to reach the half-metallic regime on the left side ferromagnet one can swap the spin valve configuration from $(\rightarrow,\uparrow)$ to $(\uparrow,\rightarrow)$. We again exploit the local pinning of the magnetization to be parallel to the $\bf d$-vector on the left side ferromagnet while the right side keeps being in the half-metallic regime. This result implies that if one ferromagnetic layer in the spin-valve is set in a magnetic configuration with maximal magnetization (i.e. $h_R>2.0)$ then by tuning the exchange field in the other ferromagnetic layer in the range [$1.-2.5$] then one can drive a series of transition from noncollinear $(\rightarrow,\uparrow)$ to collinear $(\leftarrow,\rightarrow)$ and to noncollinear $(\uparrow,\rightarrow)$ again (Fig.~\ref{f8}(c)). The changeover highlights the competition between the interface $\bf M$-$\bf d$ coupling and the $\bf M_L$-$\bf M_R$ direct exchange.

\section{Discussion and Conclusions}

We have investigated the behavior of the FSF spin valve and the most favorable magnetic configurations that can be achieved due to the presence of a spacer given by a spin-triplet superconductor. We observe that the overall phenomena can be ascribed to the occurrence of two effective couplings that control the magnetic state of the spin-valve. 
The first one is an interface coupling between the magnetic moment in the ferromagnet and the $\bf d$-vector near the superconductor-ferromagnet interface. 
The other effective interaction is due to the magnetic exchange among the magnetic moments in the ferromagnets which is directly mediated by the spin-triplet superconductor. The competition of these two interactions is shown to yield a rich variety of effects.
Indeed, we unveil how the relative strength and the interplay of these two interactions sets the magnetic ground state by exploiting the dependence on the thickness of the superconducting spacer in the spin valve.  
The interface coupling tends to align the magnetic moment in such a way that it is perpendicular or parallel to the $\bf d$-vector if the exchange field is small or large with respect to a given threshold that is close to the half-metal regime, respectively.
In contrast, the direct interaction among the magnetic moments as mediated by the spin-triplet superconductor tends to favor an anisotropic Heisenberg-like coupling which prefers an alignment of the magnetic moments in the conduction plane with a dominant antiferromagnetic character. The combination of these exchanges can set out collinear and noncollinear relative orientations of the magnetization in the ferromagnetic layers. Interestingly, the noncollinear state with perpendicularly oriented magnetic moments can be the lowest energy configuration or an excited state which is intermediate in energy with respect to the collinear ones. 

The anisotropic character of the magnetic exchange is a clear-cut signature of the spin-triplet superconductor with a single-component $\bf d$-vector. In fact, as expected by symmetry the magnetic exchange mediated by a spin-singlet superconductor is fully isotropic ~\cite{degennes1966}. Another hallmark of the spin-triplet spin-valve is that it can undergo a transition from collinear to noncollinear magnetic configurations by setting asymmetric magnetization amplitudes in the ferromagnets. This physical scenario is experimentally feasible because the amplitude of the magnetization scales with the thickness of the ferromagnet and a weaker magnetic state is obtained for thinner ferromagnets. An alternative design strategy for an asymmetric magnetic spin-valve is to employ different ferromagnetic materials with inequivalent coercive fields. 

\textcolor{black}{Regarding the magnetic anisotropy induced by the spin-triplet superconductor, we would like to stress that for the examined configuration the spin component of the triplet order parameter ($d_z$) is invariant upon rotation around the out-of-plane $z$-axis. Hence, by symmetry, there is no planar anisotropy for the magnetization which results from the spin-triplet superconductor. The physical scenario would have been different for the case of a spin-triplet superconductor with the $d$-vector lying in the $xy$ plane. For such configuration, we would have obtained a ground state with a magnetization exhibiting a planar anisotropy. These considerations indicate that the anisotropy induced by the spin-triplet Cooper pairs tends to compete with that of the ferromagnet when considering thin films with a magnetic easy axis that can vary from an out-of-plane to in-plane orientation. This is for instance the case of the itinerant ferromagnet SrRuO$_3$ where, due to the interplay of crystalline potential and spin-orbit coupling, the magnetic easy axis changes from out-of-plane to an in-plane orientation as a function of strain or thin film thickness \cite{Jung2004,Lu2013,Cuoco2022}. Such type of ferromagnets would be ideal to exploit and test the proposed effects in spin-valve hosting spin-triplet pairing.}

Since the spin-triplet superconductor spin-valve can have a ground state with a noncollinear magnetic orientation, we also expect that the critical temperature will have a nontrivial angular dependence as a function of the relative orientation of the magnetic moments in the ferromagnetic layers as compared to the case of a spin-singlet superconducting spin-valve. This behavior is completely different from that of the spin-valve based on spin-singlet superconductors which are typically designed to have a shift in the critical temperature when the magnetic orientation is changed from parallel to antiparallel in a collinear pattern. Then, superconducting spin-triplet spin-valve can act as a device to control the supercurrent with the potential to be functionalized with multiple magnetic configurations. Additionally, we notice that the magnetic mechanisms and phenomenology of the spin-valve are different from those recently obtained when considering magnetic impurities in a $p$-wave superconductor~\cite{Ouassou2023}.

It is also worth pointing out the role of spin singlet-to-spin triplet Cooper pair conversion, or viceversa, in the examined spin-valve. 
\textcolor{black}{The spin singlet-to-spin triplet Cooper pair conversion is a significant factor in superconductor-ferromagnet heterostructures and it is influenced by the inherent interface mechanisms and the noncollinear orientations of magnetization in the spin valves \cite{Bergeret2001,Valls2022, Eschrig_2015,Linder2015,Yokoyama2011}. In particular, converting spin-singlet Cooper pairs into spin-triplet pairs is essential to enable spin-triplet pairs to travel long distances and propagate effectively across superconductor-magnet heterostructures, i.e. of the type SF'FS (with F and F' having noncollinear relative magnetization). 
In contrast to spin-triplet Cooper pairs leaking into the ferromagnet, spin-singlet Cooper pairs exhibit a spatially limited proximity. This is because they are suppressed within the ferromagnet over short distances due to the pair breaking impact of the magnetic exchange.
In our study, however, we need to focus on the transformation of spin-triplet pairs to spin-singlet pairs instead of the more commonly studied conversion from singlet to triplet states.
Spin-singlet pair correlations can occur at the spin-triplet superconductor/magnet interface as a result of the breaking of translational symmetry and the existence of a magnetic exchange that breaks time-reversal symmetry.
According to our computational analysis, the amplitude of the spin-singlet pair correlations is sizable only at the F/S interface as it gets suppressed by the ferromagnetic exchange independently of the orientation of the magnetization. Hence, we expect that the triplet-to-singlet conversion can lead to contributions that are mostly affecting the interface properties with an amplitude that is however negligible in the regime of moderate-to-strong ferromagnetic exchange.
Furthermore, due to the isotropic nature of the spin-singlet correlations in the spin space, regardless of their strength or the efficiency of the triplet-to-singlet conversion process, we anticipate that they will not have a significant impact on determining the direction of the magnetization in the spin-valve.
}  

Another potentially interesting physical case for application of the phenomena examined in our work is that of a superconductor that undergoes a magnetic reconstruction on its surface~\cite{Fittipaldi2021,Mazzola2024}. In such type of superconducting configuration, we argue that the coupling between the magnetic moments at the surface and the Cooper pairs in the superconductor can lead to nonstandard magnetic response of the superconductor. For instance, apart from the reconstruction of the superconducting order parameter, the occurrence of anisotropic exchanges can also affect the time evolution of the magnetization and, thus, the magnonic excitations \cite{Poniatowski2022}. Along this line, the modification of the magnetic state on the surface of the superconductor across the superconducting transition, and the resulting spin torque induced by the coupling to the spin-triplet superconductor, can be indirectly exploited to single out the character of the electrons pairing in the superconductor. 

Finally, since the critical temperature of the superconductor can vary as a function of the thickness of the ferromagnetic leads, as shown in a recent study with a bilayer setup \cite{Hodt}, we expect a non-simple relationship between the behavior of the superconducting state and the relative thick- ness of the ferromagnets in the spin valve. A study of how the critical temperature behaves in relation to the thickness of ferromagnets in the spin valve will be the subject of future analysis.

\section*{Acknowledgments}
M.C. acknowledges support from the EU’s Horizon 2020 research and innovation program under Grant Agreement No. 964398 (SUPERGATE), from PNRR MUR project PE0000023-NQSTI, and by the QUANCOM Project 225521 (MUR PON Ricerca e Innovazione No. 2014–2020 ARS01 00734).

\textcolor{black}{
\appendix
\section{Computational procedure}}

\textcolor{black}{In this section we shortly describe the computational procedure that has been employed to obtain the solutions for the pairing amplitudes that minimize the free energy for a given orientation and strength of the magnetization in the two leads of the spin valve.
\begin{enumerate}
    \item Assuming that the size of the superconductor spin valve is $L_x$ ($L_y$) for the $x$ ($y$) sides, the decoupling of the quartic term yields a set of $4 L_x$ variational parameters $\Lambda_{i_x}=\{ \Delta^{x_+}_{i_x},\Delta^{x_-}_{i_x},\Delta^{y_+}_{i_x}, \Delta^{y_{-}}_{i_x}$\} with $i_x=1,..,L_x$, where $\Delta^{x_{\pm}}_{i_x}=\frac{1}{L_y} \sum_{i_y} \Delta_{{\bf{i,i \pm a_x}}}$ and $\Delta^{y_{\pm}}_{i_x}=\frac{1}{L_y} \sum_{i_y} \Delta_{{\bf{i,i \pm a_y}}}$. This number of variational parameters is reduced with respect to the value 2 ($L_x \times L_y$) due to the translational invariance along $y$. For the ferromagnetic subsystems, the angle $\phi$ setting the relative orientation of the magnetization in the left and right leads is given and included as an external parameter. The free energy of the $\phi$ dependent variational parameters $\Lambda_{i_x}$ is then employed to assess the energy hierarchy of the spin valve configurations. 
    \item For a given set of microscopic parameters and temperature, i.e. $\{t, \mu, V, t_{int}, h_L, h_R, T \}$, the spectrum of the bilinear mean-field Hamiltonian is obtained by standard diagonalization routines. 
    \item In order to determine the variational parameters $\Lambda_{i_x}$ at finite temperature we employ the Gibbs functional $F$, which is computed from the spectrum of the mean-field Hamiltonian $H_{MF}$ by performing the trace over all the eigenstates as $F(\Lambda_{i_x})=-\frac{1}{L_x L_y \beta} \ln(\text{Tr}\{ \exp[ -\beta H_{MF}]\}$. The parameters $\Lambda_{i_x}$ are determined by solving the coupled set of superconducting gap equations as given by $\frac{\partial F(\{\Lambda_{i_x}\})}{\partial \Lambda_{i_x}}=0$. The pairing amplitudes are calculated iteratively until the difference between successive iterations is smaller than the desired accuracy. As a starting amplitude for the iterative procedure we consider the solutions of the order parameter for the superconductor without the leads. Apart from this, we perform the analysis with several initial conditions to double check the occurrence of other solutions.
    From the solutions of the self-consistent equations we evaluate the corresponding free energy and report their behavior with regard to the exchange field, temperature and size of the superconductor. 
\end{enumerate} 
}


%


\end{document}